\documentclass[journal]{IEEEtran}
\usepackage{graphicx, subfigure}
\usepackage{float}
\usepackage{multirow}
\usepackage{amsmath}
\usepackage{cite}
\usepackage{geometry}
\usepackage{authblk}

\ifCLASSINFOpdf
\else
\fi
\begin{document}

\title{MDVSC---Wireless Model Division Video Semantic Communication}

\author[$\dagger$]{Zhicheng~Bao}
\author[$\dagger$]{Haotai~Liang}
\author[$\dagger$]{Chen~Dong \thanks{Corresponding author: dongchen@bupt.edu.cn}}
\author[$\dagger$]{Xiaodong~Xu}
\author[$\ddagger$]{Geng~Liu}
\affil[$\dagger$]{State Key Laboratory of Networking and Switching Technology, Beijing University of Posts and Telecommunications, Beijing, China}
\affil[$\ddagger$]{Smart Shine Microelectronics Technology Co.,Ltd., Beijing, China}

\maketitle


\pagestyle{empty}  
\thispagestyle{empty} 
\begin{abstract}
In this paper, we propose a new wireless video communication scheme to achieve high-efficiency video transmission over noisy channels. It exploits the idea of model division multiple access (MDMA) and extracts common semantic features across video frames. Besides, deep joint source-channel coding (JSCC) is applied to overcome the distortion caused by noisy channels. The proposed framework is collected under the name model division video semantic communication (MDVSC). In particular, temporal relative video frames are first transformed into a latent space for computing complexity reduction and data redistribution. Accordingly, a novel entropy-based variable length coding is developed further to compress semantic information under the communication bandwidth cost limitation. The whole MDVSC is an end-to-end learnable system. It can be formulated as an optimization problem whose goal is to minimize end-to-end transmission distortion under restricted communication resources. Across standard video source test sequences, test results show that the MDVSC outperforms traditional wireless video coding schemes generally under perceptual quality metrics and has the ability to control code length precisely. 
\end{abstract}


%
\IEEEpeerreviewmaketitle

\section{Introduction}
%
%
%
%
\IEEEPARstart{N}{owadays}, most wireless video communication systems' coding processes can be largely separated into source coding and channel coding. Source coding, like H.264 coding, uses many video compression algorithms to reduce the redundancies in the video sequences and then transform them into sequences of bits. Channel coding like low-density parity-check (LDPC) coding represents those sequences of bits as transmitted signals to overcome distortions caused by the imperfect wireless channel. This separated scheme greatly benefits from the decoupling of source and channel coding. 

However, the limits of the separated scheme start to emerge with more demands on higher wireless video transmission quality under lower communication resources cost. At the same time, the time-varying channel makes current wireless video transmission systems suffer from the cliff effect. When the channel condition deteriorates to a specific threshold, the video communication receiver cannot resolve the signal from the noise. Apart from that, the entropy coding relies heavily on the variational estimate of the marginal distribution of the source. Error caused by the quantization can lead to imprecise distribution estimates and finally degrade the performance of the whole video communication system.

To solve the problems mentioned above, joint source-channel coding (JSCC), which integrates source processing and channel processing together, was proposed in \cite{Fresia2010JointSA}. Within the development of deep neural networks (DNN), a new JSCC scheme based on DNNs has emerged. Works at \cite{Xie2020DeepLE, Dong2023SemanticCS} are the pioneer in the implementation of the JSCC-based semantic communication system. By using DNNs, source data can be directly encoded as continuous value symbols to be transmitted through a noisy wireless channel, thus overcoming the cliff effect and improving communication system performance. Following that, \cite{9953110} designs a deep joint source-channel coding method to achieve end-to-end video transmission over wireless channels. It is based on traditional motion estimation and motion compensation framework. Meanwhile, it exploits non-linear transform, context-driven semantic feature modeling, and rate-adaptive semantic feature transmission to outperform classical H.264/H.265 combined with LDPC and digital modulation schemes. 

However, there is still room for improvement. First, The current video semantic communication framework is based on the traditional video communication scheme, such as the motion estimation system. This system conducts some optical-flow-related operations in the pixel domain, which is highly correlated with the input dimension. For videos with high resolution and deep color depth, the time complexity for joint coding can be too high to satisfy the demand for real-time video transmission. In addition, the semantic information that humans can understand in the pixel domain may not be well encoded by the transmitter and perform video communication tasks very well. Therefore, it is the very time to design a new video communication scheme to boost the end-to-end video communication system capabilities.

Second, According to \cite{MDMA}, semantic features extracted from the same artificial intelligence (AI) model turn out to have some shared information and some personalized information. Therefore, in the context of video communication, it is assumed that semantic features extracted from the sequential video frames have much more shared information, which we name as common features. Similarly, personalized information is named individual features. In this case, common features of the entire group-of-pictures (GOP) and individual features of each video frame are the units to be transmitted instead of the traditional optical flow and residual information. That is, our system is common-individual features based and deprecates the motion estimation system. It has to be paid attention that the definition of GOP in our framework is not exactly the same as the one in the typical video coding algorithm, which is detailed in the following section. Based on the inspiration mentioned above, MDVSC is proposed. 

Besides, many variable length coding schemes in semantic communications cannot control code length manually and precisely. In \cite{9953110}, the code length turns out to be learned and controlled by the transmitter implicitly. In this case, for a given limited communication condition, such as a fixed channel bandwidth ratio, it is hard to precisely reduce the data amount while satisfying limitations. To solve that, variable length coding for secondary compression is proposed to control code length explicitly and improve the compression flexibility of the system.

To sum up, in this paper, the contribution can be summarized as the following.

(1) MDVSC Framework: We propose a new wireless video communication scheme named model division video semantic communication system (MDVSC). It exploits latent space transform and deep joint source-channel coding (JSCC) to establish an end-to-end learnable communication link.

(2) Entropy-based variable length coding: By introducing entropy as a method of measuring the importance of information, we realize variable length coding that can adjust automatically according to the threshold. 

(3) Performance validation: We verify the performance of the proposed MDVSC system across standard video source sequences. As is shown in the test results, the MDVSC can achieve much better coding gain under the low signal noise ratio (SNR) on peak signal to noise ratio (PSNR) and multiscale structural similarity (MS-SSIM) when compared with the traditional wireless video communication system. 

The rest of this paper is arranged as follows. In the next section II, we introduce the overall system model. Then, network architecture and corresponding algorithms are detailed in section III. Experiments setup and results will be shown in section IV, and section V concludes this paper. 

 




\section{The proposed method}
In this section, the system model of wireless video communication will be presented firstly. Following that, the whole framework of MDVSC will be introduced. Then, the core idea of model division video transmission and entropy-based variable length coding will be specified. In the end, the optimization goal of MDVSC will be derived.

\subsection{System model}
Assuming that there is a video sequence ${X}$ = $\{x_1, x_2, ..., x_T\}$, where the frame at time step $t$ is a vector of pixel intensities in a certain space dimension, namely $x_t \in R^m$. This sequence is divided into some GOPs, $X_g$ = $\{x_1, x_2, ..., x_N\}$ and $x_n$ is the $n$th frames in the GOP $X_g$. MDVSC takes a GOP as the input. After encoding, the GOP $X_g$ is transformed to a sequence of variable-length continuous-valued channel input symbols $\mathcal{S}_g$ = $\{s_1, s_2, ..., s_N, s_{N+1}\}$, where $s_n \in R^{k_n}$ denotes the $k_n$-dimensional channel input for frames $x_n$. To measure the cost of communication resources, channel bandwidth ratio (CBR) is proposed in \cite{Kurka2019DeepJSCCfDJ}, which is defined as CBR = $\sum_{n = 1}^{N+1} \frac{k_n}{m}$. Then the encoded GOP $\mathcal{S}_g$ is transmitted through the noisy wireless channel. The channel is modeled as a transfer function $\mathcal{W}(\cdot~;\mathcal{P})$, where $\mathcal{P}$ denotes the channel parameters. Therefore the received GOP can be formulated as $\hat{S_g}$ = $\mathcal{W}(\mathcal{S}_g;\mathcal{P})$. For the receiver, it decodes this GOP $\{\hat{s_1}, \hat{s_2}, ..., \hat{s_N}, \hat{s}_{N+1}\}$ and tries to recover $\hat{x_n}$ from the distorted $\hat{s_n}$.

It has to be made clear that the definition of group-of-pictures (GOP) in our system is only referred to a stack of video frames. While in some typical video coding algorithms, a GOP contains an intra-coded picture (I-frame of keyframe) followed by some predictive coded frames (P-frames). In our communication system, there is no concept of the I-frame or P-frame. Instead, our system takes the entire GOP into consideration.

\subsection{The framework of MDVSC}
\begin{figure*}[ht]
  \centerline{\includegraphics[width=6.8in]{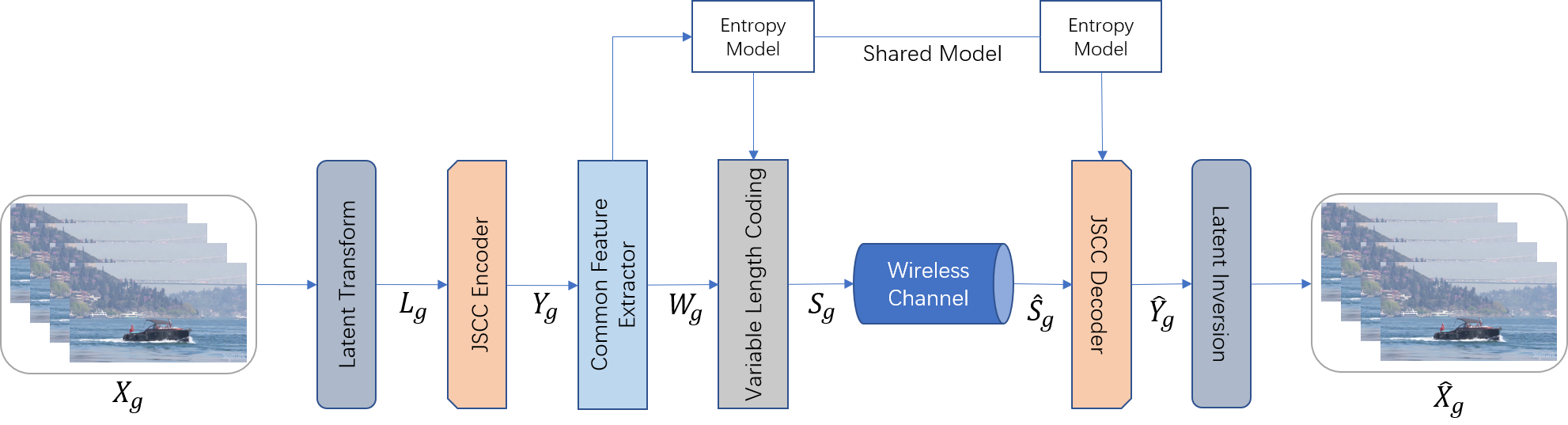}}
  \caption{The framework of MDVSC.\label{fig1}}
  \end{figure*}

The proposed MDVSC framework is illustrated in Fig.~\ref{fig1}. A stack of video frames, namely a GOP $X_g$, is the input of our system. Instead of directly encoding the input GOP, the system first transform them into latent space and form latent representation $L_g$. For one thing, it reduces the dimension of the data and accordingly speeds up computation when the spatial dimension is enormous. For another, it transforms original data distribution from visible color space, such as RGB, to a latent space. It makes it easier for the JSCC encoder to extract semantic features. 

JSCC Encoder extracts semantic features from the latent representation $L_g$ and outputs semantic feature maps $Y_g$. After that, all feature maps will be fed into the common feature extractor. It will extract common features and individual features in this GOP and generate one common feature map $W_{gc}$ along with $N$ individual feature maps $W_{gi}$ for all frames. Combining these feature maps together is collectively referred to as $W_g$. Apart from that, an entropy model is applied to restrict the common feature extractor to output valuable information. The last process at the end of the transmitter is variable length coding. Based on the $W_g$ and its entropy, some elements will be dropped to satisfy certain constraints like bandwidth cost. The output of variable length coding is denoted as $S_g$. For the existence of the shared entropy model, the receiver can reconstruct variable length coding to its original dimension.

Symbols $S_g$ are distorted to noisy signal $\hat{S_g}$ through the wireless channel. At the end of the receiver, the JSCC decoder inverses the process of the JSCC encoder and outputs latent representation $\hat{Y_g}$. Finally, the latent inversion module transforms $\hat{Y_g}$ back to its original distribution space and generates recovered GOP $\hat{X_g}$ directly. It should be noted that this framework assigns more computing burden to the transmitter, making it more practical to deploy the receiver to some computing resource-limited communication nodes.

\subsection{Model division video transmission}
\begin{figure}[ht]
  \centerline{\includegraphics[width=3in]{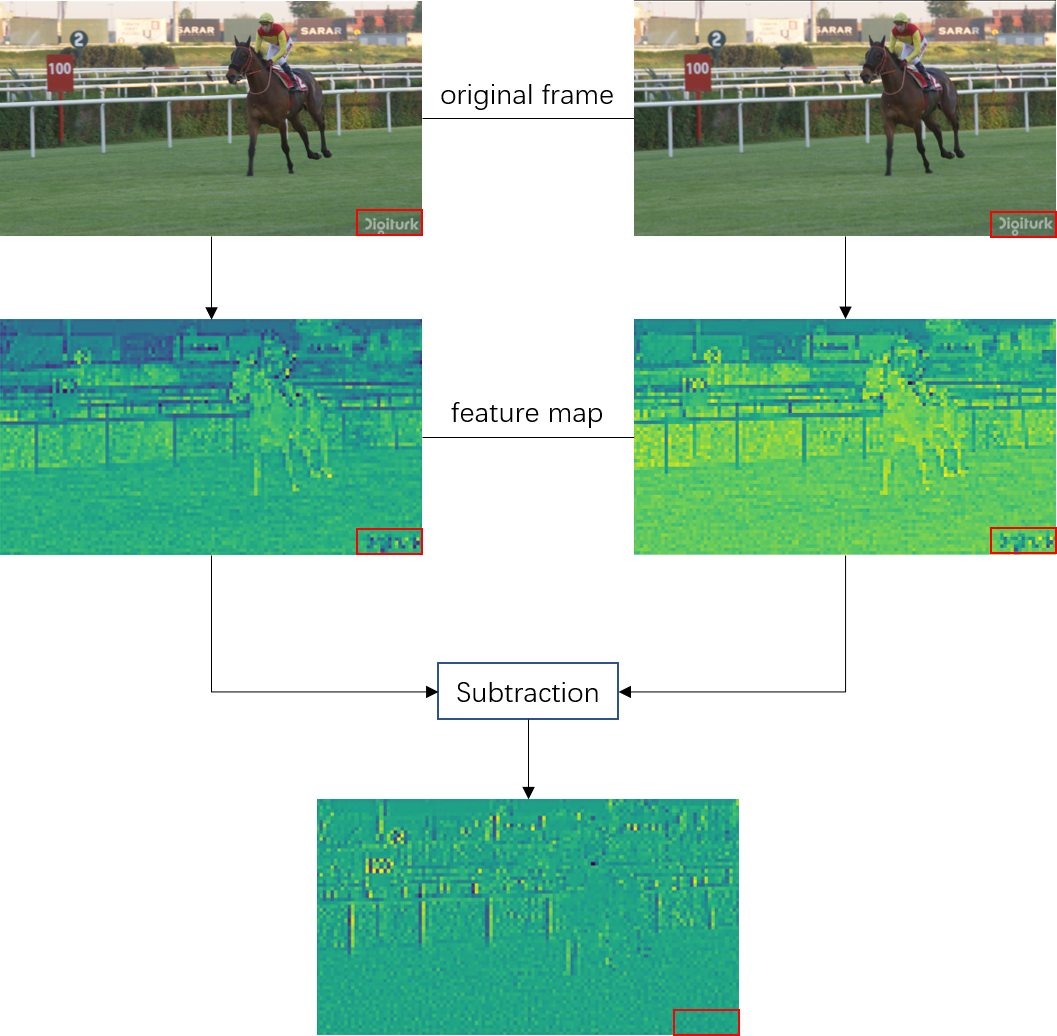}}
  \caption{The visualization of feature maps. After subtraction, common features get removed and individual features are left. It is clear that most of similar features get removed, which proves that consecutive video frames have much common features.\label{fig4}}
  \end{figure}

For consecutive video frames in a GOP, features across different frames turn out to have much shared information, which is named common features. At the same time, the left semantic information is personalized for each frame, which is named as individual features. The visualization is shown in Fig.~\ref{fig4}. Two successive video frames get encoded and transformed to feature maps. Then, subtract these two feature maps and get the output. In Fig.~\ref{fig4}, the logo's semantic information in the lower right corner of the original frame is extracted by the JSCC encoder. As it exists in all frames of this GOP, it can be viewed as a common feature of it. Therefore, for video communication tasks, it is not necessary to transmit these common features repeatedly. In fact, after subtraction, the semantic information of this logo gets removed, which means that this part of semantic information needs to be transmitted only once to save channel bandwidth. 

\begin{figure*}
  \centerline{\includegraphics[width=6in]{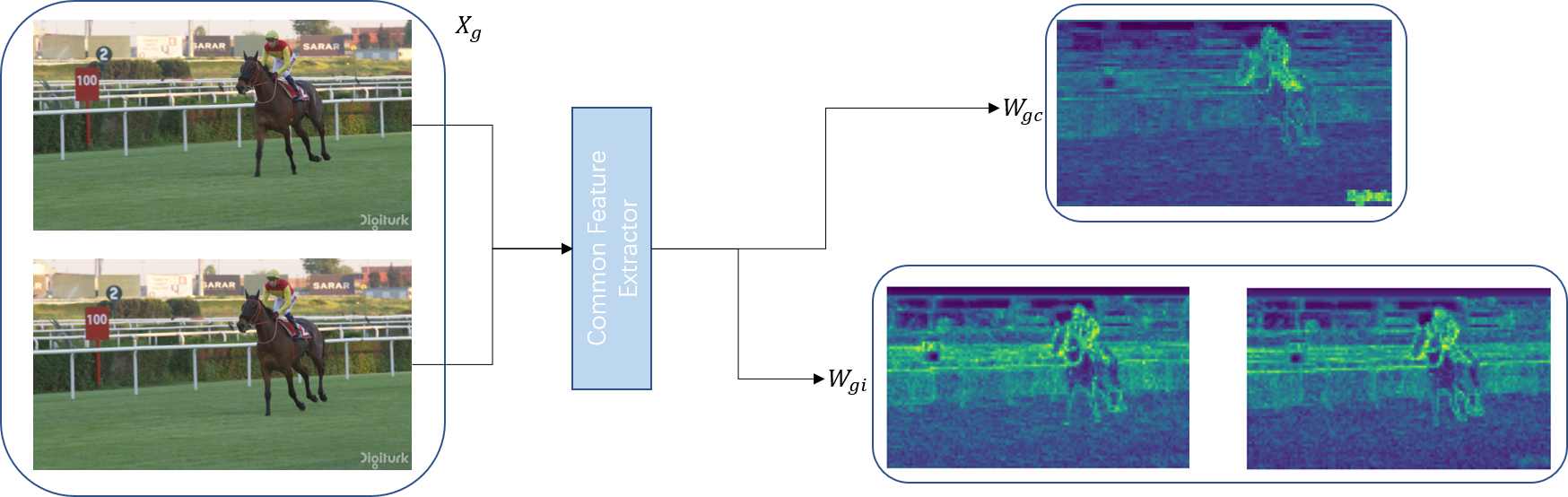}}
  \caption{Illustration of model division for common features $W_{gc}$ and individual features $W_{gi}$. Modules and encoding process before common feature extractor are emitted for simplicity.\label{fig5}}
  \end{figure*}

Based on the idea above, a model division video transmission scheme is proposed. It is presented in Fig.~\ref{fig5}. After latent transforming and JSCC encoding, the original GOP $X_g$ is fed into the common feature extractor. Common feature $W_{gc}$ represents the shared semantic information of the current GOP, while individual features $W_{gi}$ represents the personalized semantic information of each frame. As is shown in Fig.~\ref{fig5}, the extracted common feature $W_{gc}$ is more smooth and more averaged. It includes some shared information in the original GOP, such as the static logo in the lower right corner, the grassland texture, the background and the horse racer. In contrast, the individual features are sharper and more specific. They include accurate semantic information for each frame and remove shared duplicate features, such as the position of horseshoes and the outline of the horse. Meanwhile, the static logo gets removed and is not included in the individual features.

\subsection{Entropy-based variable length coding}
\begin{figure}[ht]
  \centerline{\includegraphics[width=3in]{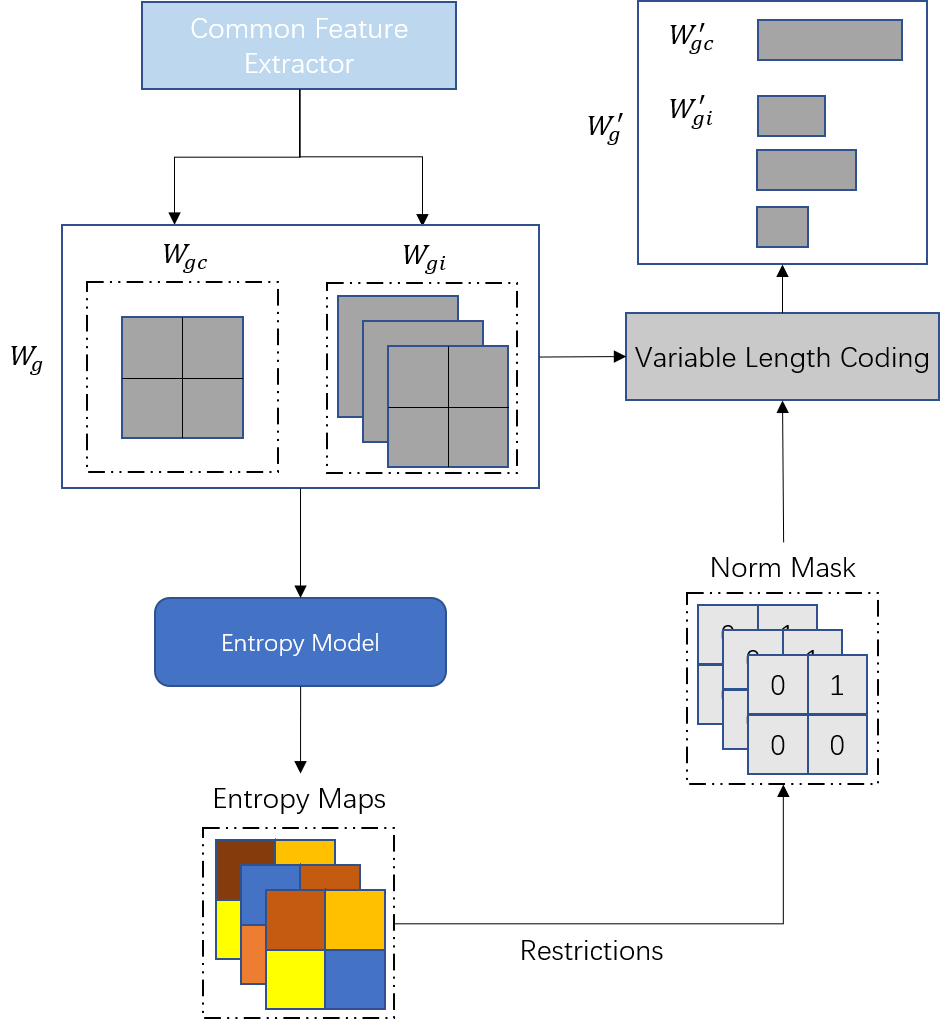}}
  \caption{Illustration of the entropy-based variable length coding scheme. Entropy is assumed to be the weight for measuring the importance of every element in feature maps.\label{fig3}}
  \end{figure}

To maximize the transmission performance while satisfying communication limitations, an explicit code length control scheme is proposed. The illustration is shown in Fig.~\ref{fig3}. Following the output $Y_g$, which is wrapped by the JSCC Encoder, the common feature extractor analyses it and outputs $W_g$. It contains a common feature map $W_{gc}$ and several individual feature maps $W_{gi}$. They will be fed into the entropy model and get entropy maps, which measure the importance of each element in $W_g$. Then, according to the restrictions, which can be set manually and explicitly, a norm mask will be calculated correspondingly. After that, the original feature maps and the norm mask can be sent into the variable length coding module and get variable length codes $W_g^{~'}$. It is composed of common feature vector $W_{gc}^{~'}$ and individual feature vectors $W_{gi}^{~'}$, both of which have controllable length. Finally, $W_g^{~'}$ gets transformed to $S_g$ and sent through the wireless channel.

\subsection{Optimization goal}
The optimization goal of the MDVSC is to improve video reconstruction quality as much as possible under limited communication resources such as CBR. Given that the MDVSC takes the GOP as the input, the loss function of the $n$th GOP can be formulated as 
\begin{equation}
    Loss = \lambda \cdot K_n + D_n = \lambda \cdot \sum_{t = 1}^{N} k_t + D_n,
\end{equation}
where $\lambda$ controls the trade-off between the CBR $K_n$ and the distortion $D_n$. Because our system learns the overall channel bandwidth allocation across sequential frames for a GOP, the $k_t$ for each frame is different. It is worth noting that in our system, the trade-off between performance and communication cost doesn't rely on the hyperparameter $\lambda$. Therefore, it is not changeable and fixed to a certain number. 

\section{Network architectures}
The details of the proposed MDVSC are illustrated in Table.~\ref{table1}. Parameter $s$ and $k$ stands for stride and kernel size respectively. Conv and Deconv represent convolution layer and transpose convolution layer. ResBlock means residual block. The whole system is asymmetric and assigns more computing burden to the transmitter. In the following subsections, details about each module will be presented.

\subsection{Transmitter}
\subsubsection{Latent Transformer}
Current video semantic communication is based on the visible space, which proposes strict requirements for the transmitter and receiver's computation capability. With the increment of video data dimension, supporting real-time video communication is becoming harder. Therefore, the latent transformer is proposed to transform video frames from the original space to the latent space. Through a down-sampling convolution layer and three successive residual layers, original data get transformed to a latent space, which is more compatible with video communications and reduces computing complexity. 

\subsubsection{JSCC Encoder}
JSCC encoder is composed of three continued blocks. There are one down-sampling convolution layer and three residual layers in each block. This module takes channel conditions into consideration and encodes latent representations after the latent transformer. Finally, channel noise robust codewords will be generated and fed into the next module. 

\subsubsection{Common Feature Extractor}
The core idea of MDVSC is to extract common and individual features in a GOP. It is achieved by a common feature extractor consisting of two successive convolution and activation layers. After outputting the common feature, the rest semantic information is the individual feature. To reduce learning difficulty of it, prior information is provided in the form of residual connection. 

\subsubsection{Entropy Model And Variable Length Coding}
The entropy model is made up of a complete auto-encoder. In some prior works, the output of the entropy model is viewed as the side information to be transmitted through the wireless channel for better frame reconstruction quality. In our system, this side information is to be transmitted for variable length decoding. Similar to \cite{Ball2018VariationalIC}, it is assumed that the probability distribution of feature maps follows a Gaussian distribution. After getting entropy maps for $W_{gc}$ and $W_{gi}$, it is sent into the variable length coding module to generate changeable length codewords. These codewords will be transmitted through the wireless channel after the power norm module.

\subsection{Channel}
The wireless channel module is a non-trainable layer that connects the transmitter and the receiver. In this paper, the additive white Gaussian noise (AWGN) channel is mainly considered. While other channel models can also be similarly incorporated by replacing this layer. The effect of wireless channel is controlled through a parameter SNR. 

\subsection{Receiver}
\subsubsection{JSCC Decoder}
Similar to the JSCC encoder, the JSCC decoder is composed of three blocks that are made up of one transpose convolution layer and three residual layers. It decodes codewords $\hat{S_g}$ to the latent representation $\hat{Y_g}$. 

\subsubsection{Latent Inversion}
At the end of the MDVSC, the latent inversion module transforms the latent representation $\hat{Y_g}$ back to the original color space, which is visible to the human. So far, the transmission of this GOP is complete. 


\begin{table}[]
\centering
\caption{Network architectures of the MDVSC}
\begin{tabular}{|c|c|}
\hline
Model                                     & Structure                        \\ \hline
Latent Transformer                        & Conv ($s$=2, $k$=5) + ResBlock * 3   \\ \hline
\multirow{3}{*}{JSCC Encoder}             & Conv ($s$=2, $k$=3) + ResBlock * 3   \\ \cline{2-2} 
                                          & Conv ($s$=2, $k$=3) + ResBlock * 3   \\ \cline{2-2} 
                                          & Conv ($s$=2, $k$=3) + ResBlock * 3   \\ \hline
\multirow{2}{*}{Common Feature Extractor} & Conv ($s$=1, $k$=3) + ReLU           \\ \cline{2-2} 
                                          & Conv ($s$=1, $k$=3) + ReLU           \\ \hline
\multirow{6}{*}{Entropy Model}            & Conv ($s$=2, $k$=3) +ReLU            \\ \cline{2-2} 
                                          & Conv ($s$=2, $k$=3) +ReLU            \\ \cline{2-2} 
                                          & Conv ($s$=2, $k$=3)                  \\ \cline{2-2} 
                                          & DeConv ($s$=2, $k$=3) + ReLU         \\ \cline{2-2} 
                                          & DeConv ($s$=2, $k$=3) + ReLU         \\ \cline{2-2} 
                                          & DeConv ($s$=2, $k$=3)                \\ \hline
\multirow{3}{*}{JSCC Decoder}             & DeConv ($s$=2, $k$=3) + ResBlock * 3 \\ \cline{2-2} 
                                          & DeConv ($s$=2, $k$=3) + ResBlock * 3 \\ \cline{2-2} 
                                          & DeConv ($s$=2, $k$=3) + ResBlock * 3 \\ \hline
Latent Inversion                          & DeConv ($s$=2, $k$=5) + ResBlock * 3 \\ \hline
\end{tabular}
\label{table1}
\end{table}

\section{Experiments}
\subsection{Experimental Setup}
\subsubsection{Datasets}
Our MDVSC is trained with the Vimeo-90k dataset \cite{Xue2017VideoEW}, which consists of 89800 video clips with a large variety of scenes and actions. Each video clip has 7 continuous frames. During training, GOP size is set to 6 and randomly cropped to 256 $\times$ 256 pixels. After training, the performance is evaluated with the HEVC test dataset \cite{Bossen2010CommonTC} and UVG dataset \cite{Mercat2020UVGD5}. Here we choose the following subsets: Class A (2560 $\times$ 1440) and UVG (3840 $\times$ 2160) for higher resolution is more usual in actual use, and it places a heavier burden on communication systems. 

\subsubsection{Implementation Details}
In all experiments, the channel dimension for latent representations and JSCC codewords is 128. The MDVSC is trained in terms of the mean square error (MSE) and tested under the PSNR, or MS-SSIM \cite{Wang2003MultiscaleSS} for perceptual quality. These three metrics can be calculated as follows: 
\begin{equation}
    MSE(X, Y) = \frac{1}{mn}\sum_{i=0}^{m-1}\sum_{j=0}^{n-1}[X(i, j)- Y(i, j)],
\end{equation}
where $m$ and $n$ denote the number of pixels horizontally and vertically.
\begin{equation}
    PSNR(X, Y) = 10 \cdot \log_{10}(\frac{1}{MSE}),
\end{equation}
\begin{equation}
\begin{split}
    MS-SSIM(X, Y) = \\
    [\mathcal{L}_M(X, Y)]^{\alpha_M}\cdot\prod_{j=1}^M[\mathcal{C}_j(X, Y)\cdot\mathcal{S}_j(X, &Y)]^{\alpha j},
\end{split}
\end{equation}
where M means different dimensions, including luminance, contrast and structure. Therefore, PSNR metric focus more on the precision of the frame reconstruction, while the MS-SSIM metric cares more about the perception of the recovered frame. The hyperparameter $\lambda$ is fixed to 8192 during all training processes as the rate-distortion (RD) tradeoff in our system does not rely on it. The learning rate for training is initialized to $10^{-4}$ and cosine annealing algorithm \cite{Loshchilov2016SGDRSG} is used for descent. A mini-batch size of 32 is used and it takes about 35 hours to train the whole MDVSC model on dual RTX A6000 GPU. 

\subsubsection{Comparison Schemes}
The MDVSC is compared with classical video coded transmission schemes including source coding H.264 \cite{Wiegand2003OverviewOT} and H.265 \cite{Sullivan2012OverviewOT}, channel coding LDPC \cite{Richardson2018DesignOL}, modulation BPSK and 16QAM for noise resistant and transmission efficiency respectively. The ffmpeg mode is set to veryslow mode. 

\subsection{Result}
\begin{figure*}[ht]
  \centerline{\includegraphics[width=6.8in]{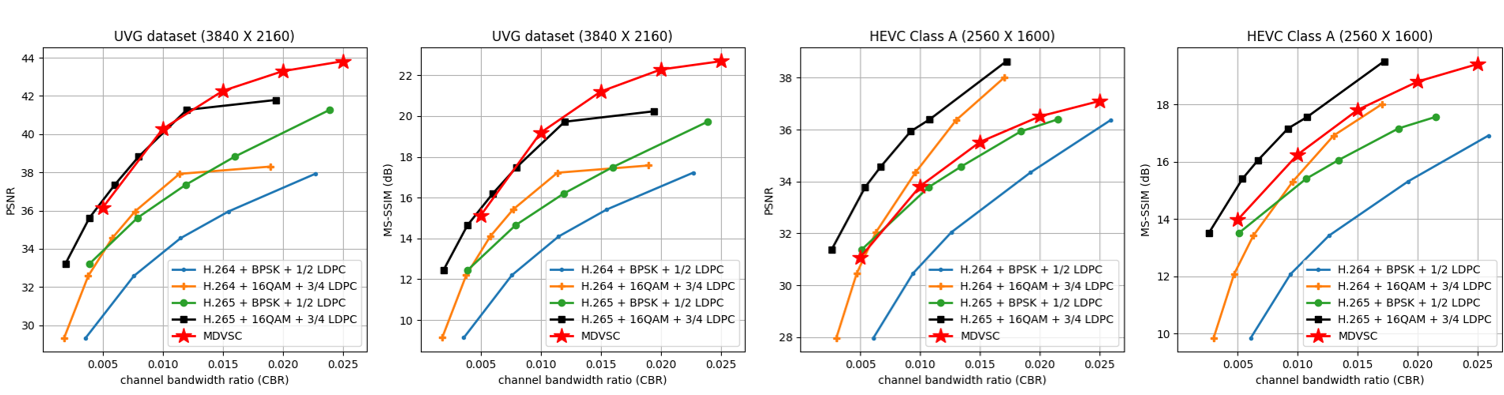}}
  \caption{PSNR and MS-SSIM performance versus the average channel bandwidth ratio (CBR) over the AWGN channel at SNR = 15dB. The value of MS-SSIM is converted using the formula -10$\log_{10}$(1 - d), where d is the MS-SSIM value in the range between zero and one. In addition, it can be found that the red star data point of the MDVSC coincides with the coordinate axis and is distributed at equal intervals even for different videos, indicating that it has precise code length control capability. While the traditional scheme can only approximate the given CBR.\label{fig6}}
  \end{figure*}

The performance curves of the PSNR and the MS-SSIM are shown in Fig.~\ref{fig6}. The SNR is set to 15dB to ensure the performance of the 16QAM. Since MS-SSIM yields values that are higher than 0.9 mostly, we convert it to dB to improve the legibility \cite{Ball2018VariationalIC}. It can be seen that under the PSNR metric, the MDVSC outperforms the noise robust combination BPSK + 1/2 LDPC generally for both H.264 and H.265. The performance of the transmission efficiency combination 16QAM + 3/4 LDPC is better than that of the MDVSC. However, under the scene of higher resolution 3840 $\times$ 2160, the MDVSC can surpass the transmission efficiency combination for H.264 and closely follow that of H.265. 

The conditions turn out to be different when under the MS-SSIM metric, which measures perceptual quality that is closer to human feelings. The MDVSC outperforms the transmission efficiency combination for H.264 and narrows the gap with that of H.265. Meanwhile, the MDVSC shows markedly coding gain as that of noise robust combination for both H.264/H.265 series. 

Besides that, it is easy to find out that the CBR parameter of the MDVSC in Fig.~\ref{fig6} is always equally spaced and ranges from 0.005 to 0.025 at the same step for different videos. It means that the MDVSC is able to control CBR precisely to meet the threshold. In contrast, other video communication schemes can only approximate the given CBR threshold.

\begin{figure}
  \centerline{\includegraphics[width=3.4in]{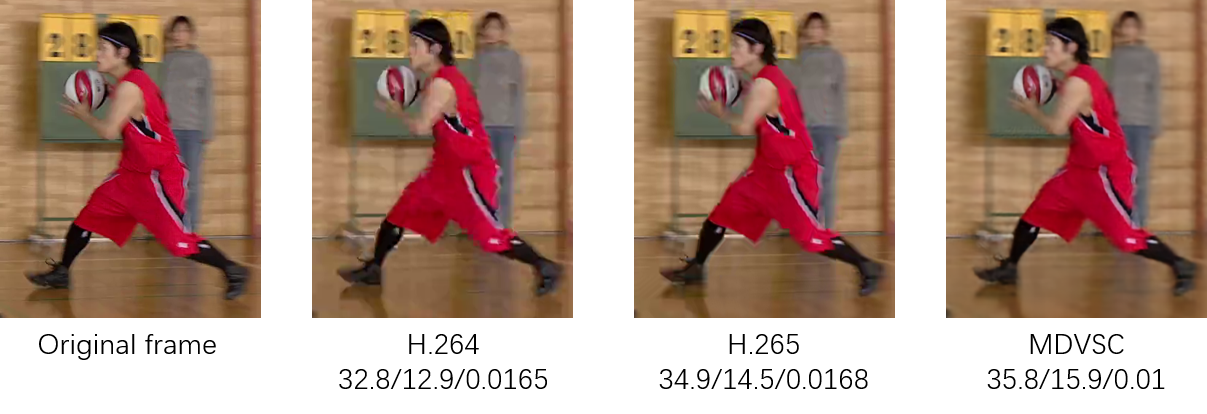}}
  \caption{Examples of visual comparison. We use videos with 1920 $\times$ 1080 pixels as lower resolution shows the difference more obviously. The subtitle is PSNR / MS-SSIM / CBR, respectively. Channel coding and modulation for H.264/H.265 are 1/2 LDPC and BPSK.\label{fig7}}
  \end{figure}
  
The visualization of the specific reconstructed frames is shown in Fig.~\ref{fig7}. One can observe that frames generated by the MDVSC are more sharp and clear, especially for the logo on the basketball and the outline of the athlete. 

\begin{figure}[ht]
  \centerline{\includegraphics[width=3.4in]{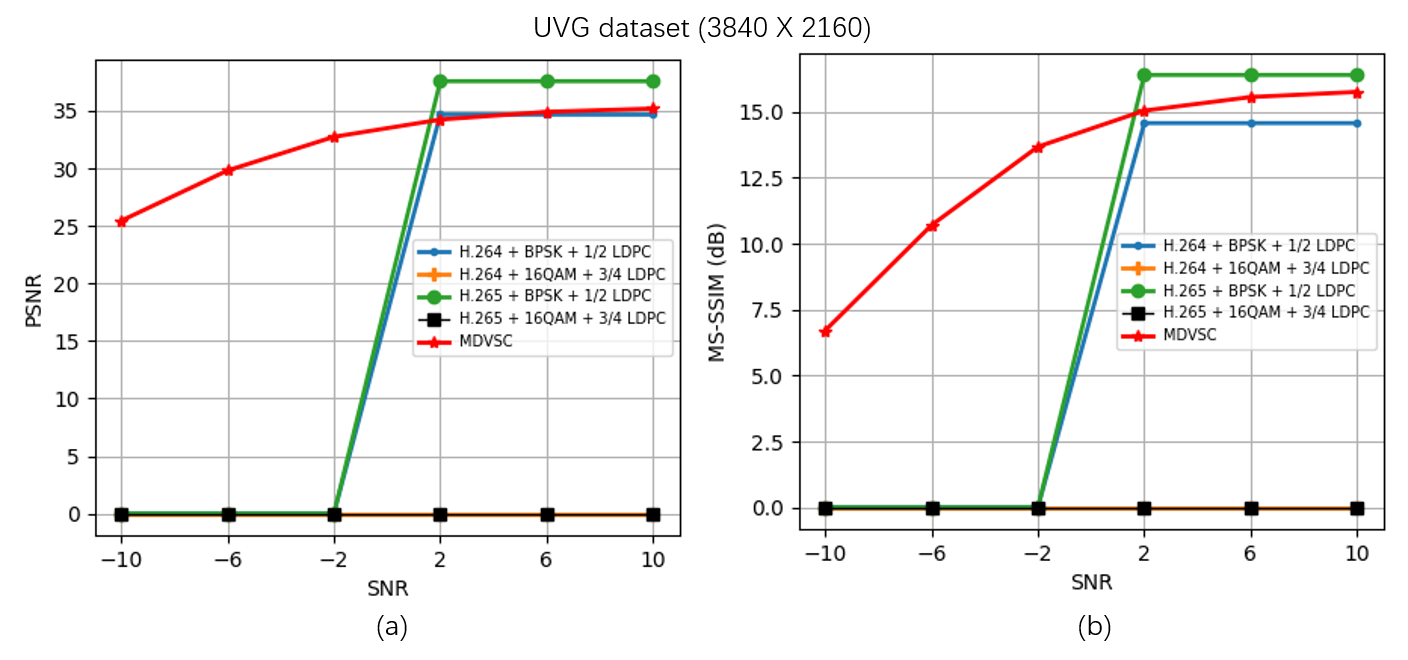}}
  \caption{PSNR and MS-SSIM performance versus the change of channel SNR over AWGN channel. \label{fig8}}
  \end{figure}

Fig.~\ref{fig8} also provides the PSNR and MS-SSIM results versus the change of channel SNR, where the CBR constraint is set to 0.01. It has to be paid attention that the MDVSC is trained only once under the SNR 10dB and tested under variable SNR. There is no model fine-tuning for different target SNR. It can be observed that the MDVSC performs well even under extremely severe channel conditions. In contrast, traditional schemes suffer from significant cliff-effect and cannot work properly. 

Besides, in the case of downlink video transmission, the decoding time at the receiver is tested. For the video with 3840 $\times$ 2160 resolution, we use a 13900K CPU and RTX A6000 GPU to run the decoder. Test results show that the average decoding time for one frame is about 8ms, making it possible to decode 4K video in real time for semantic communications.

\section{Conclusion}
In this paper, a new deep JSCC-based model division video semantic communication scheme over wireless channels is proposed. It was collected under the name MDVSC. This MDVSC framework exploits latent space transformation for better coding gain, exploits model division feature extraction for higher communication efficiency, and finally utilizes entropy-based variable length coding for explicit communication resource cost control. Experiment results have shown that the proposed MDVSC can generally surpass traditional wireless video communication schemes under low SNR and have precise code length control capability.


%





\ifCLASSOPTIONcaptionsoff
  \newpage
\fi



\bibliographystyle{IEEEtran}
%



\bibliography{IEEEabrv, reference}



%








\end{document}